\documentclass[
 aip,
 amsmath,amssymb,
 groupedaddress,
 reprint
]{revtex4-1}

\usepackage{graphicx}% Include figure files
\usepackage{dcolumn}% Align table columns on decimal point
\usepackage{bm}% bold math
\usepackage[utf8]{inputenc}
\usepackage[T1]{fontenc}
\usepackage{mathptmx}
\usepackage{etoolbox}

\usepackage{braket}
\usepackage{color}

\begin{document}

\preprint{AIP/123-QED}

\title{Fast parametric two-qubit gate for highly detuned fixed-frequency superconducting qubits using a double-transmon coupler}
\author{Kentaro Kubo}
\email[Author to whom correspondence should be addressed: Kentaro Kubo, ]{kentaro3.kubo@toshiba.co.jp}
\affiliation{ 
Frontier Research Laboratory, Corporate Research$\&$Development Center, Toshiba Corporation, 1, Komukai-Toshiba-cho, Saiwai-ku, Kawasaki 212-8582, Japan.
}
\author{Hayato Goto}
%\email[\gt{Hayato Goto, }]{\gt{hayato1.goto@toshiba.co.jp}}
\affiliation{ 
Frontier Research Laboratory, Corporate Research$\&$Development Center, Toshiba Corporation, 1, Komukai-Toshiba-cho, Saiwai-ku, Kawasaki 212-8582, Japan.
}
%\email[\kb{Authors to whom correspondence should be addressed: Kentaro Kubo, }]{\kb{kentaro3.kubo@toshiba.co.jp}}
%Authors to whom correspondence should be addressed: 
%[Kentaro Kubo, kentaro3.kubo@toshiba.co.jp; 
%Hayato Goto, hayato1.goto@toshiba.co.jp]

\date{\today}

\begin{abstract}
High-performance two-qubit gates have been reported 
with superconducting qubits coupled via a single-transmon coupler (STC). 
Most of them are implemented for qubits with a small detuning 
since reducing residual $ZZ$ coupling for highly detuned qubits by an STC is challenging. 
In terms of the frequency crowding and crosstalk, 
however, 
highly detuned qubits are desirable. 
Here, we numerically demonstrate a high-performance parametric gate 
for highly detuned fixed-frequency qubits 
using 
a recently proposed tunable coupler called a double-transmon coupler (DTC). 
Applying an ac flux pulse, 
we can perform a maximally entangling 
universal gate ($\sqrt{\rm iSWAP}$) 
with an average fidelity over 99.99$\%$ and a short gate time of about 24 ns. 
This speed is comparable to resonance-based gates for slightly detuned tunable qubits. 
Moreover, using a dc flux pulse alternatively, we can achieve 
another kind of entangling gate called a CZ gate with an average fidelity over 99.99$\%$ and a gate time of about 18 ns. 
Given the flexibility and feasible settings, 
we can expect that the DTC will contribute to realizing 
a high-performance quantum computer in the near future.

\end{abstract}

\maketitle

Recently, 
tunable couplers have become a key component in superconducting quantum computers. 
By tuning external parameters, such as magnetic flux, 
it can turn on and off the coupling between computational qubits
, leading to high-performance two-qubit gates.

Various types of tunable couplers have been proposed before, 
including 
the inductive coupler
\cite{inductive1_parametric1,inductive2_parametric2,inductive3,inductive4,inductive5,inductive7}, 
flux-qubit coupler\cite{fluxqubit1,fluxqubit2}, 
fluxonium coupler\cite{fluxonium1, fluxonium2}, 
transmon coupler\cite{yan2018tunable,STCChinese,IBMSTCZZ, IBMSTC,PRXSTC,RiggettiSTC,APLtunabletransmon,spremacy1,spremacy2,
goto2022DTC}, 
and so on. 
In particular, 
the one containing a frequency-tunable transmon %in itself 
has been widely studied, because of its high performance
\cite{yan2018tunable,STCChinese,IBMSTCZZ,IBMSTC,PRXSTC,RiggettiSTC,spremacy1,spremacy2,APLtunabletransmon}. 
We refer to this type of coupler as a single-transmon coupler (STC), 
in order to distinguish it from the double-transmon coupler (DTC)\cite{goto2022DTC} described below.
The effective coupling strength between two qubits coupled via an STC
is determined by competing of direct coupling strength between qubits 
and virtual coupling strength via a coupler transmon\cite{yan2018tunable}. 
Several groups have reported fast and accurate two-qubit gate implementations using an STC\cite{spremacy1,spremacy2,STCChinese,IBMSTCZZ, IBMSTC,PRXSTC,RiggettiSTC}.
For example, 
a CZ gate with a 99.85$\%$ gate fidelity and a $46$ ns gate time has been implemented for fixed-frequency transmons by applying a dc magnetic flux into an STC\cite{IBMSTC}. 
Entangling gates comparable to this example 
have been implemented using a resonance-based method for tunable qubits\cite{PRXSTC,RiggettiSTC}. 
Moreover, 
the STC was also used to demonstrate quantum advantage\cite{spremacy1,spremacy2}.

However, 
it is known that reducing residual $ZZ$ coupling for highly detuned qubits is challenging for the STC\cite{STCChinese, IBMSTCZZ, IBMSTC,goto2022DTC}. 
For example, 
it has been reported that there are about 20 kHz and 85 kHz residual $ZZ$ couplings
for about 650 MHz and 360 MHz detuned qubits, respectively\cite{STCChinese, IBMSTCZZ}. 
Because of this drawback, 
the STC is often used for slightly detuned qubits (less than qubit anharmonicity)\cite{PRXSTC,RiggettiSTC} with a few kHz residual $ZZ$ coupling. 
In these cases, 
we should pay attention to issues of qubit-frequency crowding and crosstalk between qubits.

To compensate this drawback, 
the DTC have recently been proposed\cite{goto2022DTC}. 
As its name suggests, 
the DTC is a tunable coupler containing two transmons in itself 
and follows a different operating principle from the STC. 
It allows us to easily reduce the residual $ZZ$ coupling strength for highly detuned qubits\cite{goto2022DTC}. 
It has also been reported that 
a CZ gate with a high fidelity over 99.99$\%$ and a short gate time of about 24 ns can be implemented 
for fixed-frequency qubits with 0.7 GHz detuing by applying a dc flux pulse to the DTC\cite{goto2022DTC}. 
Thus, the DTC is an alternative approach to a
high-performance coupler.

As one of the typical methods to implement a two-qubit gate, 
parametric gates have been actively studied with and without a tunable coupler
\cite{inductive1_parametric1,inductive2_parametric2,parametric3,parametric4, IBMparametric, riggetti_parametric1, riggetti_parametric2, riggetti_parametric3, parametric_noguchi}. 
For example, 
in Ref.\ \onlinecite{IBMparametric}, 
by modulating the tunable-transmon frequency in STC 
at the two-qubit frequency difference (854 MHz),
parametric exchange coupling was implemented for highly detuned fixed-frequency qubits. 
Because of such an operating principle, 
parametric gates are suitable for highly detuned qubits. 
However, 
its gate speed is very slow, 
i.e., iSWAP gates with a 183 ns gate time for fixed-frequency qubits\cite{IBMparametric} 
and with a 175 ns gate times for tunable qubits\cite{parametric3}. 
Due to this drawback, 
for fast gate implementations, 
it is common to use resonance-based methods\cite{PRXSTC,RiggettiSTC} 
instead of the parametric coupling method.
In resonance-based methods, 
the tunability of computational-qubit frequencies is indispensable. 
Therefore, this method is undesirable with respect to flux-noise sensitivity 
compared to the case of fixed-frequency qubits.
Moreover, 
it is obviously incompatible with highly detuned qubits and thus weak for frequency crowding and crosstalk. 
We aim to overcome these drawback by the parametric-coupling method using the DTC.

In this work, 
we show that by only applying an ac magnetic flux to the DTC, 
a parametric gate (including a $\sqrt{\rm iSWAP}$) for highly detuned qubits
with speed and fidelity comparable to the resonance-based gates\cite{PRXSTC,RiggettiSTC} 
can be implemented without qubit-frequency tuning. 
Furthermore, 
high-fidelity CPHASE gate (including a CZ)\cite{goto2022DTC} 
can also be implemented in the same setup as the parametric gate only by using a dc flux pulse, 
instead of the ac one. 
Combining these high-performance parametric and CPHASE 
gates makes gate depths shallower, 
thus implementing efficient quantum computation in a near-term machine
\cite{NISQ1, NISQ6,NISQ9}.
These results indicate that the DTC is a highly promising element to realize a practical quantum computer.

\begin{figure}[t]
  \includegraphics[width=0.5\textwidth]{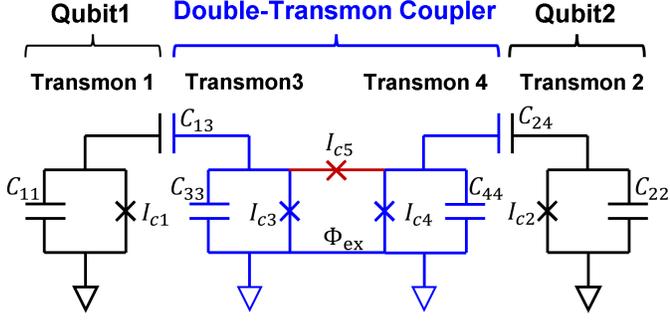}
  \caption{
  	Diagram of the double-transmon coupler with two fixed-frequency qubits.
  }
  \label{fig:design}
\end{figure}

The DTC consists of two fixed-frequency transmons (Transmons 3 and 4) coupled through 
a common loop including an adittional Josephson junction, 
as shown in blue and red in Fig.\ {\ref{fig:design}}\cite{goto2022DTC}. 
Two computational qubits [fixed-frequency transmons (Transmons 1 and 2) shown in black in Fig.\ {\ref{fig:design}}] are capacitively coupled to the DTC. 
$C_{ij}\ (i,j \in \{1, 2, 3, 4\})$ is capacitance between the $i$th and $j$th transmons, 
$I_{ci}\ (i \in \{1, 2, 3, 4, 5\})$ is the critical current of the $i$th Josephson junction, 
and $\Phi_{\rm ex}$ is the external magnetic flux in the loop of the DTC. 

The Hamiltonian of this system is written as\cite{goto2022DTC}
\begin{align}
	\hat{H} &= 4\hbar\hat{\mathbf{n}}^{T} W \hat{\mathbf{n}} 
	+ \hbar \frac{\dot{\Theta}_{\rm ex}}{\omega_{C_{34}}}
	\begin{pmatrix}
	  0 & 0 & -1 & 1 \\
	\end{pmatrix}
	W\hat{\mathbf{n}} + \hat{V}, 
	\label{eq:Hamiltonian}\\
	\hat{V} &= -\sum_{i=1}^4 \hbar\omega_{J_i}\cos\hat{\varphi}_i
		-\hbar\omega_{J5}\cos(\hat{\varphi}_{4}-\hat{\varphi}_3-\Theta_{\rm ex}), 
\end{align}
where 
$\hbar$ is the reduced Planck constant, 
$\Theta_{\rm ex}=\Phi_{\rm ex}/\phi_0$ is an angle defined with $\Phi_{\rm ex}$, 
$\hbar W = e^2M^{-1}/2$ 
with a capacitor matrix $M$ ($M_{ii}=\sum_{j=1}^4C_{ij}$ and $M_{ij}=-C_{ij}$ for $i\neq j$), 
and $\hbar \omega_{C_{34}}=e^2/(2C_{34})$ with the elementary charge $e$. 
Operators $\hat{n}_i$,  $\hat{\varphi}_{i}$, and $\hbar\omega_{J_i}=\phi_0 I_{ci}$ are, respectively, 
the Cooper-pair number operator, 
the phase difference operator, 
and the Josephson energy for the $i$th Josephson junction. 
Operators $\hat{n}_i$ and $\hat{\varphi}_{i}$ satisfy 
the canonical commutation relation $[\hat{\varphi}_{i}, \hat{n}_j]=i\delta_{i,j}$. 

The matrix representations of the operators 
in the basis of $\hat{n}_i$ eigenfuctions
are given as follows: 
\begin{align}
	\hat{n}_i=
		\begin{pmatrix}
			  -N & \ & \ \\
 			  \ &\ddots &\ \\ 
 			  \ &\ &N \\
		\end{pmatrix}, 
	\cos\hat{\varphi}_i=\frac{1}{2}
		\begin{pmatrix}
			  \ &1 &\ &\ \\
 			  1 &\ &\ddots &\ \\ 
 			  \ &\ddots &\ &1 \\
 			  \ &\ &1 &\  \\
		\end{pmatrix}, 
\end{align}
where $N$ is a cutoff for the Cooper-pair number. 
Each single-transmon operator is expressed as a $(2N+1)\times(2N+1)$ matrix. 
In this work, 
we choose $N=10$ for sufficient convergence of energies 
and perform all numerical calculations using the Quantum Toolbox in Python (QuTiP)
\cite{qutip1,qutip2}.

Parameter values used in this work are shown in Table\ \ref{tab:parameters}. 
We take transmon frequencies $\omega_i$ and capacitance $C_{ij}$ as design values. 
Since transmons in this frequency range 
with coherence times of several 10 $\mathrm{\mu}$s
have been widely used
\cite{feasibility1,feasibility2,feasibility3}, 
our setting is experimentally feasible. 
The other parameters 
are calculated using the design values.  
By definition, $W_{ij}$ is uniquely determined from $C_{ij}$. 
The Josephson frequencies of Transmons 1-4 $\omega_{Ji}$ are calculated as 
\begin{align}
	\omega_{Ji} = \frac{\left(\omega_{i}+W_{ii}\right)^2}{8W_{ii}}. 
\end{align}
As for $I_{c5}(\omega_{J5})$, 
we set $r_J$, 
the ratio of $I_{c5}$ to the average value of $I_{c3}$ and $I_{c4}$, 
to 0.3. 
It is higher than $r_J=0.25$ used in the previous work\cite{goto2022DTC}. 
This setting is preferable to achieve for both fast parametric and CPHASE gates, as explained later.

\begin{table}[t]
	{\small
  		\begin{minipage}[htbp]{0.45\columnwidth}
  			\vspace{-5.42cm}
  			\centering
  			\begin{tabular}{|m{9.7em}|m{3em}|} \hline
    				$\omega_{1}/2\pi$ (GHz) & {\bf 7.0} \\ \hline
    				$\omega_{2}/2\pi$ (GHz)  & {\bf 7.7} \\ \hline
   				$\omega_{3}/2\pi$ (GHz) & {\bf 10.2} \\ \hline
  		  		$\omega_{4}/2\pi$ (GHz) & {\bf 10.2} \\ \hline
  			\end{tabular}
  		
	  		\vspace{0.3cm}
	  		
	  		  \begin{tabular}{|m{9.7em}|m{3em}|} \hline
  	 			$\omega_{J1}/2\pi$ (GHz) & 22.5 \\ \hline
   			 	$\omega_{J2}/2\pi$ (GHz)  & 27.0 \\ \hline
   				$\omega_{J3}/2\pi$ (GHz) & 47.2 \\ \hline
    				$\omega_{J4}/2\pi$ (GHz) & 47.2 \\ \hline
   			 	$\omega_{J5}/2\pi$ (GHz) & 14.2 \\ \hline
  			\end{tabular}
			
			\vspace{0.3cm}
			
		     \begin{tabular}{|m{9.7em}|m{3em}|} \hline
		     	$I_{c1}$ (nA) & 24.4 \\ \hline
		    		$I_{c2}$ (nA)  & 31.3 \\ \hline
		    		$I_{c3}$ (nA) & 53.6 \\ \hline
		    		$I_{c4}$ (nA) & 53.6 \\ \hline
		    		$I_{c5}$ (nA) &  17.9 \\ \hline
			\end{tabular}
			
			  \vspace{0.3cm}
		     \begin{tabular}{|m{9.7em}|m{3em}|} \hline
		     	$r_{J}=I_{c5}/\left[\left(I_{c3}+I_{c4}\right)/2\right]$ & 0.3 \\ \hline
			\end{tabular}

  \end{minipage}
  }
  \hspace{0.04\columnwidth}
  {\small
  		\begin{minipage}[b]{0.45\columnwidth}
  			\centering
  			\begin{tabular}{|m{8.5em}|m{3.5em}|} \hline
    				$C_{11}$ (fF) & {\bf 60.0} \\ \hline
				$C_{12}$ (fF)  & {\bf 0.025} \\ \hline
				$C_{13}$ (fF) & {\bf 6.0} \\ \hline
 			   	$C_{14}$ (fF) & {\bf 0.05} \\ \hline
 			   	$C_{22}$ (fF)  & {\bf 60.0} \\ \hline
			   	$C_{23}$ (fF) & {\bf 0.05} \\ \hline
			    	$C_{24}$ (fF) & {\bf 6.0} \\ \hline
			    	$C_{33}$ (fF) & {\bf 60.0} \\ \hline
 			   	$C_{34}$ (fF) & {\bf 1.0} \\ \hline
			    	$C_{44}$ (fF) & {\bf 60.0} \\ \hline
			\end{tabular}	

			\vspace{0.3cm}
			
			\begin{tabular}{|m{8.5em}|m{3.5em}|} \hline
    				$W_{11}/2\pi$ (MHz) & 296 \\ \hline
    				$W_{12}/2\pi$ (MHz)  & 0.189 \\ \hline
   				$W_{13}/2\pi$ (MHz) & 26.5 \\ \hline
   				$W_{14}/2\pi$ (MHz) & 0.632 \\ \hline
    				$W_{22}/2\pi$ (MHz)  & 296 \\ \hline
   				$W_{23}/2\pi$ (MHz) & 0.632 \\ \hline
   				$W_{24}/2\pi$ (MHz) & 26.5 \\ \hline
    				$W_{33}/2\pi$ (MHz) & 291 \\ \hline
    				$W_{34}/2\pi$ (MHz) & 4.42 \\ \hline
    				$W_{44}/2\pi$ (MHz) & 291 \\ \hline
  			\end{tabular}
  			
  			%\begin{tabular}{|m{7em}|m{5em}|} \hline
  	 		%	$g_{12}/2\pi$(MHz) & 2.44 \\ \hline
  	 		%	$g_{13}/2\pi$(MHz) & 394 \\ \hline
  	 		%	$g_{14}/2\pi$(MHz) & 9.42 \\ \hline
  	 		%	$g_{23}/2\pi$(MHz) & 9.86 \\ \hline
  	 		%	$g_{24}/2\pi$(MHz) & 413 \\ \hline
  	 		%	$g_{34}/2\pi$(MHz) & 79.6 \\ \hline
  			%\end{tabular}
  \end{minipage}
  }
\caption{
	Parameter setting in this work. 
	Bold values are design values. 
	The others are calculated using the design values. }
\label{tab:parameters}
\end{table}

\begin{figure*}[htbp]
  \includegraphics[width=\textwidth]{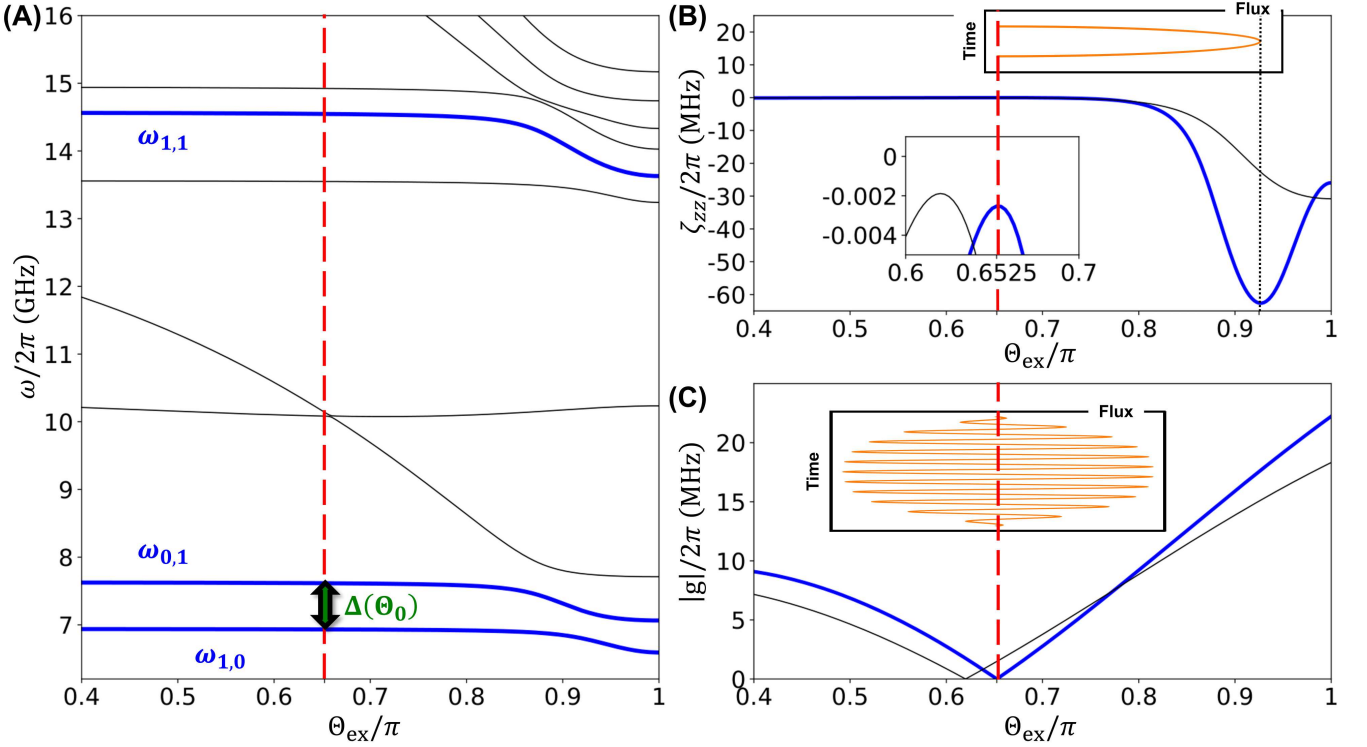}
  \caption{
  	(A) Eigenfrequencies as functions of $\Theta_{\rm ex}$. 
  		Bold lines are 
  		$\omega_{1,0}(\Theta_{\rm ex}), \omega_{0,1}(\Theta_{\rm ex})$ 
		and $\omega_{1,1}(\Theta_{\rm ex})$, 
		from bottom. 
		The vertical dashed lines indicate the idling point $\Theta_{\rm ex}=\Theta_{0}$. 
  	(B) $ZZ$ coupling strength as a function of $\Theta_{\rm ex}$. 
  		Bold and thin curves are $\zeta_{ZZ}(\Theta_{\rm ex})$ for $r_{J}=0.3$ and $0.25$, respectively. 
  		The middle inset shows enlarged graph of $\Theta_{\rm ex} \in [0.6,0.7]$. 
  		The vertical dotted line indicate the value of $\Theta_{\rm ex}$ 
  		where $|\zeta_{ZZ}|$ takes maximum value. 
  		The right top inset shows an conceptual image of the dc flux pulse for the CPHASE gate. 
  	(C) Effective transverse coupling strength as a function of $\Theta_{\rm ex}$.
  		Bold and thin curves are $|g(\Theta_{\rm ex})|$ for $r_{J}=0.3$ and $r_{J}=0.25$, respectively. 
  		The inset shows an conceptual image of the ac flux pulse for a parametric gate. 
  }
  \label{fig:E}
\end{figure*}

Eigenfrequencies  
as functions of $\Theta_{\rm ex}$ are shown in Fig.\ \ref{fig:E}(A). 
Bold lines in Fig.\ \ref{fig:E}(A) show 
$\omega_{1,0}(\Theta_{\rm ex}), \omega_{0,1}(\Theta_{\rm ex})$,  
and $\omega_{1,1}(\Theta_{\rm ex})$, 
from bottom,  
where $\omega_{i,j}$ denotes the eigenfrequency of the two-qubit state $\ket{ij}$ 
and $\omega_{0,0}$ is set to the origin. 
Thin lines are eigenfrequencies of other states, 
such as the coupler-transmon excited states and the qubit second excited states.

The $ZZ$ coupling strength $\zeta_{ZZ}$ is defined as 
\begin{align}
	\zeta_{ZZ}(\Theta_{\rm ex})
	=\omega_{1,1}(\Theta_{\rm ex})
	-\left[\omega_{1,0}(\Theta_{\rm ex})+\omega_{0,1}(\Theta_{\rm ex})\right]
	, 
\end{align}
and its $\Theta_{\rm ex}$ dependence is shown in Fig.\ \ref{fig:E}(B). 
Bold and thin curves are, respectively,  
$\zeta_{ZZ}(\Theta_{\rm ex})$ for $r_{J}=0.3$ and 
$0.25$ (for comparison, which is used in the previous work\cite{goto2022DTC}). 
Compared with $\zeta_{ZZ}(\Theta_{\rm ex})$ for $r_{J}=0.25$,
the behavior of the one for $r_{J}=0.3$ 
changes from a monotonic decrease to a dip structure. 
This dip structure increases 
the maximum value of $|\zeta_{ZZ}|$, 
and therefore it is possible to achieve a faster CPHASE gate by applying a dc flux pulse like the right top inset of Fig.\ \ref{fig:E}(B), 
as explained later. 

In contrast to the previous result\cite{goto2022DTC}, 
there is no rigourous zero point of $\zeta_{ZZ}$, 
as shown in the middle inset of Fig.\ \ref{fig:E}(B). 
This can be regarded at the cost of parameter settings 
that allow us to achieve fast parametric and CPHASE gates.  
However, the minimum value of $|\zeta_{ZZ}|/(2\pi)\simeq2.53\ {\rm kHz}$ 
at $\Theta_{\rm ex}=\Theta_{0}\equiv0.6525\pi$ 
is still small enough to use practically. 
In fact, 
even when rigorous zero points exist, 
it is difficult to avoid this degree of residual $ZZ$ coupling in acutual experiments 
(for example, see 
Fig.\ 14 in Ref.\ \onlinecite{PRXSTC}). 
Even in such cases\cite{IBMSTC,PRXSTC} 
implementation of two-qubit gates with a fidelity higher than 99$\%$ was reported. 
Therefore, we also tolerate this residual $ZZ$ coupling, 
and thus we define the qubit idling states $\ket{00}, \ket{10},\ket{01},\ket{11}$ 
by the energy eigenstates at $\Theta_{\rm ex}=\Theta_{0}$. 
All vertical and horizontal dashed lines in this paper describe this idling point. 
\begin{figure*}[htbp]
	\includegraphics[width=\textwidth]{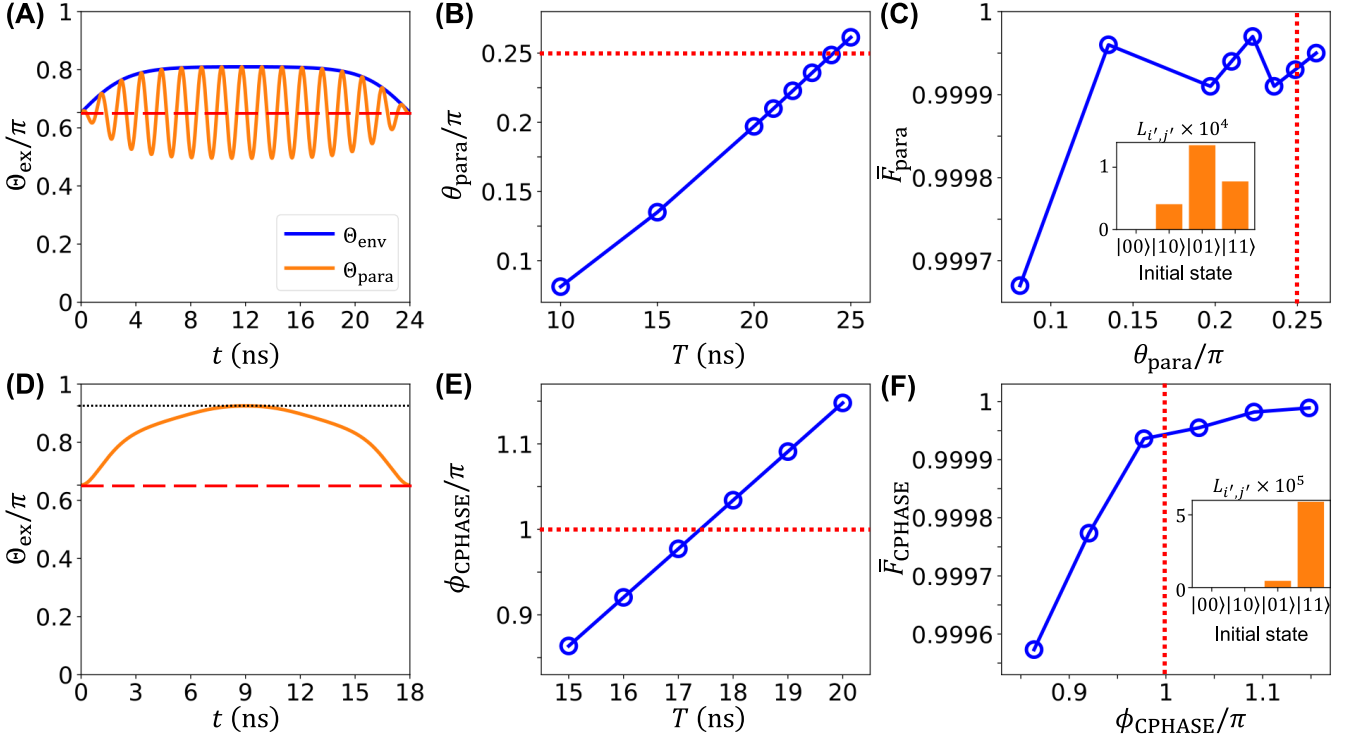}
 	\caption{
		(A)-(C) The parametric gate performance with the double-transmon coupler. 
  		(A) Ac gate pulse shape of $\Theta_{\rm ex}$ in the case of $T=24\ {\rm ns}$. 
  			 It realize 
  			 maximally entangling universal gate ($\sqrt{\rm iSWAP}$). 
 		(B) Rotation angle $\theta_{\rm para}$ of 
 			the parametric gate as a function of gate time $T$. 
 		(C) Average parametric-gate fidelity $\bar{F}_{\rm para}$ 
 			as a function of rotation angle $\theta_{\rm para}$. 
 			The inset shows initial-state dependence of leakage error rates 
 			at the end of the $\sqrt{\rm iSWAP}$ gate. 
		In (B) and (C), dotted lines indicate $\theta_{\rm para}=\pi/4$
		: the angle of $\sqrt{\rm iSWAP}$ gate.
 		(D)-(F) The CPHASE gate performance with the double-transmon coupler. 
  		(D) Dc gate pulse shape of $\Theta_{\rm ex}$ in the case of $T=18\ {\rm ns}$. 
  			It realize CZ gate.
  		(E) The phase angle $\phi_{\rm CPHASE}$ of 
  			the CPHASE gate as a function of gate time $T$. 
 		(F) Average CPHASE-gate fidelity $\bar{F}_{\rm CPHASE}$ 
 		 	as a function of the rotation angle $\phi_{\rm CPHASE}$. 
 		 	The inset shows initial-state dependence of leakage error rates 
 				at the end of the CZ gate. 
 		 In (D) and (F), dotted lines indicate 
 		 $\phi_{\rm CPHASE}=\pi$, correspond to the CZ gate.
 		 }
  	\label{fig:gate}
\end{figure*}

In a region where 
two-qubit detuning
$\Delta(\Theta_{\rm ex})\equiv \omega_{0,1}(\Theta_{\rm ex})-\omega_{1,0}(\Theta_{\rm ex})$ can be approximated to $\Delta(\Theta_{0})$, 
the two-qubit Hamiltonian in an appropriate rotating frame is well approximated as follows: 
\begin{align}
	\hat{H}_{\rm eff}(\Theta_{\rm ex}) = \hbar g(\Theta_{\rm ex}) e^{-i \Delta(\Theta_{0}) t/\hbar}\ket{01}\bra{10}+{\rm h.c.},
	\label{eq:Heff}
\end{align}
where $g(\Theta_{\rm ex})=\bra{01}\hat{H}\ket{10}/\hbar$ is the effective transverse coupling strength between two computational qubits. 
Note that, by definition, $g(\Theta_{\rm ex})$ is exactly zero at the idling point $\Theta_{\rm ex}=\Theta_0$.
Its $\Theta_{\rm ex}$ dependence is shown in Fig.\ \ref{fig:E}(C). 
Bold and thin curves are 
$|g(\Theta_{\rm ex})|$ for $r_{J}=0.3$ and 
$0.25$ (for comparison), respectively. 
As shown by these curves, 
$|g(\Theta_{\rm ex})|$ almost linearly varies around zero. 
Therefore, 
modulating $\Theta_{\rm ex}$ at a frequency of $\Delta(\Theta_{0})$ 
leads to 
a parametric gate that rotates the probability amplitudes 
of $\ket{01}$ and $\ket{10}$. 
In this work, 
we set $\Delta(\Theta_0)/(2\pi)$ to about 700 MHz 
(see Table\ \ref{tab:parameters}). 
As suggested by Fig.\ \ref{fig:E}(C), 
this two-qubit detuning will be much larger than 
transverse coupling strength induced by flux noise. 
Hence, 
it can be expected that the proposed parametric gate is robust against flux noise.

To implement the above parametric gate, 
we introduce the following ac flux pulse.
\begin{align}
	\Theta_{\rm para}(t)&=\Theta_{0}+\Theta_{\rm env}(t)\cos\left[\Delta(\Theta_{0})t\right], \label{eq:pulse}\\
	\Theta_{\rm env}(t)&=\alpha\tanh(\beta t)\tanh\left[\beta (T-t)\right], 
\end{align}
where $\Theta_{\rm env}$ is the pulse envelope with parameters $\alpha$ and $\beta$. 
A conceptual diagram of this pulse is shown in the inset of Fig.\ \ref{fig:E}(C). 
The larger the amplitude 
of $g(\Theta_{\rm ex})$ is, 
the faster the parametric gate is.
However, 
if $\alpha$ is too large, 
$\Delta(\Theta_{\rm para})$ deviates so much from $\Delta(\Theta_0)$ that 
Eq.\ (\ref{eq:Heff}) does not hold and hence a parametric gate does not work well. 
To avoid this problem and achieve a high-fidelity parametric gate, 
we set $\alpha$ to $0.1575\pi$ such that $\max_{t}\Theta_{\rm para}(t)$ is $0.81\pi$.  
As for $\beta$, 
it should be 
small enough to suppress nonadiabatic errors. 
We thus set $\beta=0.3\ \mathrm{GHz}$. 
The ac flux pulse with these parameters and a gate time 
$T=24\ {\rm ns}$ is shown in Fig.\ \ref{fig:gate}(A).
This ac flux pulse realizes a fast and accurate maximally entangling universal gate ($\sqrt{\rm iSWAP}$), as shown later.

To evaluate gate performance, 
we calculate an average gate fidelity.
In the case of two-qubit gates, 
the formula for an average fidelity $\bar{F}$ is given by\cite{kueng2016comparing,goto2022DTC} 
\begin{align}
	\bar{F}
		=\frac{\left|{\rm tr}(\hat{U}_{\rm id}^{\dag}\hat{U}')\right|^2
		+{\rm tr}\left(\hat{U}'^{\dag}\hat{U}'\right)}{20}, 
		\label{eq:fid}
\end{align}
where $\hat{U}_{\rm id}$ is an ideal gate operation matrix 
and $\hat{U}'$ is the matrix determined by simulation results, as explained below. 
In the case of the present parametric gate, 
$\hat{U}_{\rm id}$ is given by the following form. 
\begin{align}
	\hat{U}_{\rm para}(\theta_{\rm para})
		=
		\begin{pmatrix}
			  1 & 0 & 0 & 0 \\
 			  0 & e^{i\phi_{11}}\cos\theta_{\rm para} & -ie^{i\phi_{12}}\sin\theta_{\rm para} & 0 \\
 			  0 & -ie^{i\phi_{21}}\sin\theta_{\rm para} & e^{i\phi_{22}}\cos\theta_{\rm para} & 0 \\
 			  0 & 0 & 0 & e^{i(\phi_{11}+\phi_{22})} \\
		\end{pmatrix},  
\end{align}
where $e^{i\phi_{21}}=e^{i(\phi_{11}+\phi_{22}-\phi_{12})}$ due to unitarity. 
$\hat{U}_{\rm para}(\pi/4)$ is a maximally entangling gate
and equivalent to the $\sqrt{\rm iSWAP}$ gate up to single-qubit phase rotations. 
We aim to implement this gate. 

The matrix 
$\hat{U}'$ is determined by the numerical result of a gate operation. 
Using $\widetilde{\ket{i'j'}}$, 
which is the resultant state of a gate operation to $\ket{i'j'}$, 
$\hat{U}'$ is defined as 
\begin{align}
	U'_{2i+j,2i'+j'}
	=\frac{\braket{00|\widetilde{00}}^*}{\left|\braket{00|\widetilde{00}}\right|}
	 \times \braket{ij|\widetilde{i'j'}}. 
\end{align}
Here, 
we choose an overall phase factor, 
$\braket{00|\widetilde{00}}^*/\left|\braket{00|\widetilde{00}}\right|$, 
such that 
the phase factor of $U'_{0,0}$ is equal to 1. 
Using $\hat{U}'$, 
we define parameter of $\hat{U}_{\rm para}$ as 
$\theta_{\rm para}=\arcsin|U'_{1,2}|$, 
$e^{i\phi_{ii}}=U'_{i,i}/|U'_{i,i}|\ (i=1,2)$, and 
$e^{i\phi_{12}}=iU'_{1,2}/|U'_{1,2}|$.
We calculate the average fidelity $\bar{F}_{\rm para}$ from Eq.\ (\ref{eq:fid}) using these $\hat{U}_{\rm id}$ and $\hat{U}'$.

Here, note that $\hat{U}'$ is not unitary in general because of leakage errors, which 
are the main cause of the gate infidelity. 
In the case where the initial state is $\ket{i'j'}$, 
the leakage error rate $L_{i',j'}$ is calculated from $\hat{U}'$ as
\begin{align}
	L_{i',j'}=1-\sum_{i,j}\left|U'_{2i+j,2i'+j'}\right|^2. 
\end{align}
Suppressing $L_{i',j'}$ by tuning the pulse form 
is crucial for achieving a high-fidelity gate.

Figure\ \ref{fig:gate}(B) shows that the rotation angle of the parametric gate $\theta_{\rm para}$ increases almost linearly with the gate time $T$. 
In particular, 
the $\sqrt{\rm iSWAP}$ gate corresponding to $\theta_{\rm para}=\pi/4$ 
is achieved at $T\simeq24\ \rm{ns}$, 
as indicated by the horizontal dotted line in Fig.\ \ref{fig:gate}(B). 
The initial-state dependence of the leakage error rates 
at the end of this gate 
is shown in the inset of Fig.\ \ref{fig:gate}(C). 
Since the detunings between $\omega_{0,1}(\Theta_{\rm ex})$, 
or $\omega_{1,1}(\Theta_{\rm ex})$, 
and frequencies of coupler-transmon excited states 
are close to the modulation frequency $\Delta(\Theta_0)$ 
around $\Theta_{\rm ex}=\max_{t}\Theta_{\rm para}(t)=0.81\pi$ [see Fig.\ \ref{fig:E}(A)], 
$L_{0,1}$ and $L_{1,1}$ are relatively larger than others. 
However, they are small enough to achieve a high-fidelity gate. 
In fact, 
the average fidelity of this $\sqrt{\rm iSWAP}$ gate, 
indicated by the vertical dotted line in Fig.\ \ref{fig:gate}(C), 
is higher than $99.99\%$. 
Thus, 
the DTC with an ac flux pulse allows us to achieve a fast and high-fidelity parametric gate. 
This speed is comparable to resonance-based gates\cite{RiggettiSTC, PRXSTC} 
and 
also much faster than previously reported parametric gates 
for highly detuned fixed-frequency qubits 
\cite{inductive1_parametric1,inductive2_parametric2,IBMparametric}.

Next, we show that the CPHASE gate can also be implemented by just changing the waveform of the flux pulse. 
Implementing a fast CPHASE gate requires large $|\zeta_{ZZ}|$.
Therefore, we introduce a dc flux pulse $\Theta_{\rm CPHASE}$ 
with a maximum value indicated by the dotted line in Fig.\ \ref{fig:E}(B),  
where $|\zeta_{ZZ}|$ is maximized 
[see the conceptual diagram in the right top inset of Fig.\ \ref{fig:E}(B)].
The details of the waveform are tuned by the technique known to suppress leakage errors\cite{goto2022DTC,martinis2014fast}.
Figure\ \ref{fig:gate}(D) shows a dc flux pulse optimized for $T=18\ {\rm ns}$. 
This dc flux pulse realizes a fast and accurate CZ gate, as shown later. 

In the case of the CPHASE gate, 
$\hat{U}_{\rm id}$ is given as follows: 
\begin{align}
\hat{U}_{\rm CPHASE}={\rm diag}(1,e^{i\phi_{11}},e^{i\phi_{22}},e^{i\phi_{33}}).
\end{align}
From $\hat{U}'$, 
we define $e^{i\phi_{ii}}$ as $e^{i\phi_{ii}}=U'_{i,i}/|U'_{i,i}|$ at $i=1,2,3$
and phase angle of CPHASE gate as 
$\phi_{\rm CPHASE}=\phi_{33}-\phi_{22}-\phi_{11}$ . 
Figure \ref{fig:gate}(D) shows 
that $\phi_{\rm CPHASE}$ increases almost linearly with the gate time $T$. 
In particular, 
the CZ gate, 
corresponding to $\phi_{\rm CPHASE}=\pi$, 
is achieved when $T\simeq18\ \rm{ns}$, 
as indicated by the horizontal dotted line in Fig.\ \ref{fig:gate}(E). 
Leakage error rates  
are negligibly small,  
as shown in the inset of Fig. \ref{fig:gate}(F). 
The average fidelity of the CZ gate, 
indicated by the vertical dotted line in Fig.\ \ref{fig:gate}(F), 
is higher than $99.99\%$. 
Remarkably, this gate is faster than the CZ gate with 24 ns demonstrated in the previous work\cite{goto2022DTC}.

Some readers might think that the $\sqrt{\rm iSWAP}$ gate with an AC flux pulse is not needed 
since it can be implemented by two CZ and multiple single-qubit gates\cite{decomp}. 
However, 
since this composite gate takes at least about $36$ ns ($18\times 2$ ns), 
the $\sqrt{\rm iSWAP}$ gate with an AC flux pulse is much faster. 
Therefore, the ability to implement both $\sqrt{\rm iSWAP}$ and CZ gates independently is an advantage of our proposal.

In conclusion, 
we have demonstrated that a high-performance $\sqrt{\rm iSWAP}$ gate 
can be implemented for highly detuned fixed-frequency qubits 
by parametric coupling with a DTC, where the flux in the loop of the DTC is modulated at a frequency of qubit detunig. 
This gate is as fast as resonance-based gates\cite{PRXSTC,RiggettiSTC}. 
Moreover, 
high-performance CZ gate can also be implemented in the same setup 
as the parametric gate 
only by using a dc flux pulse, instead of the ac one. 
Our implementation works well in an experimentally 
feasible setting shown in Table\ \ref{tab:parameters}. 
Because of the large qubit detuning, 
it can be expected to resist to frequency crowding and qubit crosstalk. 
The ability to implement two kinds of high-performance gates ($\sqrt{\rm iSWAP}$ and CZ) 
only by changing the flux pulse form leads to shallower gate depths
, resulting in efficient quantum computation with 
NISQ devices\cite{NISQ1, NISQ6,NISQ9}. 
These results indicate that the DTC is highly promising for a near-term practical quantum computer.

%\section*{Conflict of interest}
%\vspace{-4mm}
%H.G. is an inventor on a Japanese patent application related to this work filed by the Toshiba Corporation (no. P2019-164742, filed 30 July 2021). The authors declare that they have no other competing interests.

%\section*{Author Contributions}

%\section*{Data Availability}
%\vspace{-4mm}
%The data that support the findings of this study are available
%from the corresponding author upon reasonable request.

%reference
%\bibliographystyle{unsrt} 

\end{document}